\documentclass[twocolumn]{aastex631}

\usepackage{graphicx}%

\newcommand{\JWST}{\textit{JWST}}

\begin{document}

\title{The First Billion Years, According to JWST}

\author{Participants of the ISSI Breakthrough Workshop 2024}
\affiliation{International Space Science Institute, Hallerstrasse 6, 3012 Bern, Switzerland }

\author[0000-0002-8192-8091]{Angela Adamo}
\affiliation{Department of Astronomy, The Oskar Klein Centre, Stockholm University, AlbaNova, SE-10691 Stockholm, Sweden}

\author[0000-0002-7570-0824]{Hakim Atek}
\affiliation{Institut d'Astrophysique de Paris, CNRS, Sorbonne Universit\'e, 98bis Boulevard Arago, 75014, Paris, France}

\author[0000-0002-9921-9218]{Micaela B. Bagley}
\affiliation{Department of Astronomy, The University of Texas at Austin, 2515 Speedway, Stop C1400, Austin, TX 78712, USA}

\author[0000-0002-2931-7824]{Eduardo Ba{\~n}ados}
\affiliation{Max Planck Institut f\"ur Astronomie, K\"onigstuhl 17, D-69117, Heidelberg, Germany}

\author[0000-0002-8638-1697]{Kirk S.~S.~Barrow}
\affiliation{Department of Astronomy, University of Illinois at Urbana-Champaign, 
1002 W Green St, Urbana, IL 61801, USA}

\author[0000-0002-4153-053X]{Danielle A. Berg}
\affiliation{Department of Astronomy, The University of Texas at Austin, 2515 Speedway, Stop C1400, Austin, TX 78712, USA}

\author[0000-0001-5063-8254]{Rachel Bezanson}
\affiliation{Department of Physics and Astronomy and PITT PACC, University of Pittsburgh, Pittsburgh, PA 15260, USA}

\author[0000-0001-5984-0395]{Maru\v{s}a Brada{\v c}}
\affiliation{University of Ljubljana, Faculty of Mathematics and Physics, Jadranska ulica 19, SI-1000 Ljubljana, Slovenia}
\affiliation{Department of Physics and Astronomy, University of California Davis, 1 Shields Avenue, Davis, CA 95616, USA}

\author[0000-0003-2680-005X]{Gabriel Brammer}
\affiliation{Cosmic Dawn Center (DAWN)}
\affiliation{Niels Bohr Institute, University of Copenhagen, Jagtvej 128, DK-2200, Copenhagen N, Denmark}

\author[0000-0002-1482-5818]{Adam C. Carnall}
\affiliation{Institute for Astronomy, University of Edinburgh, Royal Observatory, Edinburgh EH9 3HJ, UK}

\author[0000-0002-0302-2577]{John Chisholm}
\affiliation{Department of Astronomy, The University of Texas at Austin, 2515 Speedway, Stop C1400, Austin, TX 78712, USA}

\author[0000-0001-7410-7669]{Dan Coe}
\affiliation{Space Telescope Science Institute, 3700 San Martin Drive, Baltimore, MD 21218, USA}
\affiliation{Association of Universities for Research in Astronomy (AURA), Inc.~for the European Space Agency (ESA)}
\affiliation{Department of Physics and Astronomy, The Johns Hopkins University, 3400 N Charles St. Baltimore, MD 21218, USA}

\author[0000-0001-8460-1564]{Pratika Dayal}
\affiliation{Kapteyn Astronomical Institute, University of Groningen, 9700 AV Groningen, The Netherlands}

\author[0000-0002-2929-3121]{Daniel J.\ Eisenstein}
\affiliation{Center for Astrophysics $|$ Harvard \& Smithsonian, 60 Garden St., Cambridge MA 02138 USA}

\author[0000-0002-1722-6343]{Jan J. Eldridge}
\affiliation{Department of Physics, University of Auckland, Private Bag 92019, Auckland, New Zealand}

\author[0000-0002-9400-7312]{Andrea Ferrara}
\affiliation{Scuola Normale Superiore,  Piazza dei Cavalieri 7, 50126 Pisa, Italy}

\author[0000-0001-7201-5066]{Seiji Fujimoto}
\affiliation{Department of Astronomy, The University of Texas at Austin, Austin, TX 78712, USA}

\author[0000-0002-2380-9801]{Anna de Graaff}\affiliation{Max-Planck-Institut f\"ur Astronomie, K\"onigstuhl 17, D-69117, Heidelberg, Germany}

\author[0000-0003-4750-0187]{Melanie Habouzit}\affiliation{Max-Planck-Institut f\"ur Astronomie, K\"onigstuhl 17, D-69117, Heidelberg, Germany}

\author[0000-0001-6251-4988]{Taylor A.\ Hutchison}	
\altaffiliation{NASA Postdoctoral Fellow}
\affiliation{Astrophysics Science Division, Code 660, NASA Goddard Space Flight Center, 8800 Greenbelt Rd., Greenbelt, MD 20771, USA}

\author[0000-0001-9187-3605]{Jeyhan S. Kartaltepe}
\affiliation{Laboratory for Multiwavelength Astrophysics, School of Physics and Astronomy, Rochester Institute of Technology, 84 Lomb Memorial Drive, Rochester, NY 14623, USA}

\author[0000-0002-3838-8093]{Susan A. Kassin}
\affiliation{Space Telescope Science Institute, 3700 San Martin Drive, Baltimore, MD 21210, USA}
\affiliation{Department of Physics \& Astronomy, Johns Hopkins University, 3400 N. Charles Street, Baltimore, MD 21218, USA}

\author[0000-0002-7613-9872]{Mariska Kriek}\affiliation{Leiden Observatory, Leiden University, P.O. Box 9513, 2300 RA Leiden, The Netherlands}

\author[0000-0002-2057-5376]{Ivo Labb\'e}
\affiliation{Centre for Astrophysics and Supercomputing, Swinburne University of Technology, Melbourne, VIC 3122, Australia}

\author[0000-0002-4985-3819]{Roberto Maiolino}
\affiliation{Kavli Institute for Cosmology, University of Cambridge, Madingley Road, Cambridge, CB3 OHA, UK}

\affiliation{Cavendish Laboratory - Astrophysics Group, University of Cambridge, 19 JJ Thomson Avenue, Cambridge, CB3 OHE, UK}

\affiliation{Department of Physics and Astronomy, University College London, Gower Street, London WC1E 6BT, UK}

\author[0000-0001-8442-1846]{Rui Marques-Chaves}
\affiliation{Department of Astronomy, University of Geneva, Chemin Pegasi 51, 1290 Versoix, Switzerland}

\author[0000-0003-0695-4414]{Michael V. Maseda}
\affiliation{Department of Astronomy, University of Wisconsin-Madison, 475 N. Charter St., Madison, WI 53706, USA}

\author[0000-0002-3407-1785]{Charlotte Mason}
\affiliation{Cosmic Dawn Center (DAWN)}
\affiliation{Niels Bohr Institute, University of Copenhagen, Jagtvej 128, DK-2200, Copenhagen N, Denmark}

\author[0000-0003-2871-127X]{Jorryt Matthee}
\affiliation{Institute of Science and Technology Austria (ISTA), Am Campus 1, 3400 Klosterneuburg, Austria}

\author[0000-0001-5538-2614]{Kristen B. W. McQuinn}
\affiliation{Department of Physics and Astronomy, Rutgers, the State University of New Jersey,  136 Frelinghuysen Road, Piscataway, NJ 08854, USA}
\affiliation{Space Telescope Science Institute, 3700 San Martin Drive, Baltimore, MD, 21218, USA}

\author[0000-0001-6181-1323]{Georges Meynet}
\affiliation{Department of Astronomy, University of Geneva, Chemin Pegasi 51, 1290 Versoix, Switzerland}

\author[0000-0003-3997-5705]{Rohan P. Naidu}
\altaffiliation{NASA Hubble Fellow}
\affiliation{MIT Kavli Institute for Astrophysics and Space Research, 77 Massachusetts Ave., Cambridge, MA 02139, USA}

\author[0000-0001-5851-6649]{Pascal A. Oesch}
\affiliation{Department of Astronomy, University of Geneva, Chemin Pegasi 51, 1290 Versoix, Switzerland}
\affiliation{Cosmic Dawn Center (DAWN)}
\affiliation{Niels Bohr Institute, University of Copenhagen, Jagtvej 128, DK-2200, Copenhagen N, Denmark}

\author[0000-0001-8940-6768]{Laura Pentericci}
\affiliation{INAF --OAR Osservatorio Astronomico di Roma, via di Frascati 33 00078 Monte Porzio Catone, Italy}

\author[0000-0003-4528-5639]{Pablo G. P\'erez-Gonz\'alez}
\affiliation{Centro de Astrobiolog\'{\i}a (CAB), CSIC-INTA, Ctra. de Ajalvir km 4, Torrej\'on de Ardoz, E-28850, Madrid, Spain}

\author[0000-0002-7627-6551]{Jane R. Rigby}
\affiliation{Astrophysics Science Division, Code 660, NASA Goddard Space Flight Center, 8800 Greenbelt Rd., Greenbelt, MD 20771, USA}

\author[0000-0002-4140-1367]{Guido Roberts-Borsani}
\affiliation{Department of Astronomy, University of Geneva, Chemin Pegasi 51, 1290 Versoix, Switzerland}

\author[0000-0001-7144-7182]{Daniel Schaerer}
\affiliation{Department of Astronomy, University of Geneva, Chemin Pegasi 51, 1290 Versoix, Switzerland}

\author[0000-0003-3509-4855]{Alice E. Shapley}
\affiliation{Department of Physics and Astronomy, University of California, Los Angeles, 430 Portola Plaza, Los Angeles, CA 90095}

\author[0000-0001-6106-5172]{Daniel P. Stark}\affiliation{Department of Astronomy, University of Arizona, 933 N Cherry Avenue, Tucson, AZ 85721}

\author[0000-0001-9935-6047]{Massimo Stiavelli}
\affiliation{Space Telescope Science Institute, 3700 San Martin Drive, Baltimore, MD 21218, USA}

\author[0000-0001-6369-1636]{Allison L. Strom}
\affiliation{Center for Interdisciplinary Exploration and Research in Astrophysics (CIERA), Northwestern University, 1800 Sherman Ave., Evanston, IL, 60201, USA}
\affiliation{Department of Physics and Astronomy, Northwestern University, 2145 Sheridan Road, Evanston, IL 60208, USA}

\author[0000-0002-5057-135X]{Eros Vanzella}
\affiliation{INAF -- OAS, Osservatorio di Astrofisica e Scienza dello Spazio di Bologna, via Gobetti 93/3, I-40129 Bologna, Italy}

\author[0000-0002-7633-431X]{Feige Wang}
\affiliation{Steward Observatory, University of Arizona, 933 N Cherry Avenue, Tucson, AZ 85721, USA}

\author[0000-0003-3903-6935]{Stephen M.~Wilkins} %
\affiliation{Astronomy Centre, University of Sussex, Falmer, Brighton BN1 9QH, UK}
\affiliation{Institute of Space Sciences and Astronomy, University of Malta, Msida MSD 2080, Malta}

\author[0000-0003-2919-7495]{Christina C. Williams}
\affiliation{NSF’s National Optical-Infrared Astronomy Research Laboratory, 950 North Cherry Avenue, Tucson, AZ 85719, USA}

\author[0000-0002-4201-7367]{Chris J. Willott}
\affiliation{NRC Herzberg, 5071 West Saanich Rd, Victoria, BC V9E 2E7, Canada}

\author[0000-0003-2212-6045]{Dominika Wylezalek}
\affiliation{Astronomisches Rechen-Institut, Zentrum für Astronomie der Universität Heidelberg, Mönchhofstr. 12-14, D-69120 Heidelberg, Germany}

\author[0009-0007-8087-6975]{Antonella Nota}
\affiliation{International Space Science Institute, Hallerstrasse 6, 3012 Bern, Switzerland }
\affiliation{Space Telescope Science Institute, 3700 San Martin Drive, Baltimore, MD 21218, USA}


\correspondingauthor{Antonella Nota}\email{antonella.nota@issibern.ch}







\begin{abstract}
With stunning clarity, \JWST\ has revealed the 
Universe's first billion years. The scientific community is analyzing a wealth of \JWST\ imaging and spectroscopic data from that era, and is in the process of rewriting the astronomy textbooks. Here, 1.5 years into the \JWST\ science mission, we provide a snapshot of the great progress made towards understanding the initial chapters of our cosmic history.
We highlight discoveries and breakthroughs, topics and issues that are not yet understood, and questions that will be addressed in the coming years, as \JWST\ continues its revolutionary observations of the Early Universe. While this compendium is written by a small number of authors, invited to ISSI Bern in March 2024 as part of the 2024 ISSI Breakthrough Workshop, we acknowledge the work of a large community that is advancing our collective understanding of the evolution of the Early Universe.
\end{abstract}


\section{Introduction} \label{sec:intro}

\begin{figure*}[h!]
    \centering
    \includegraphics[width=0.9\textwidth]{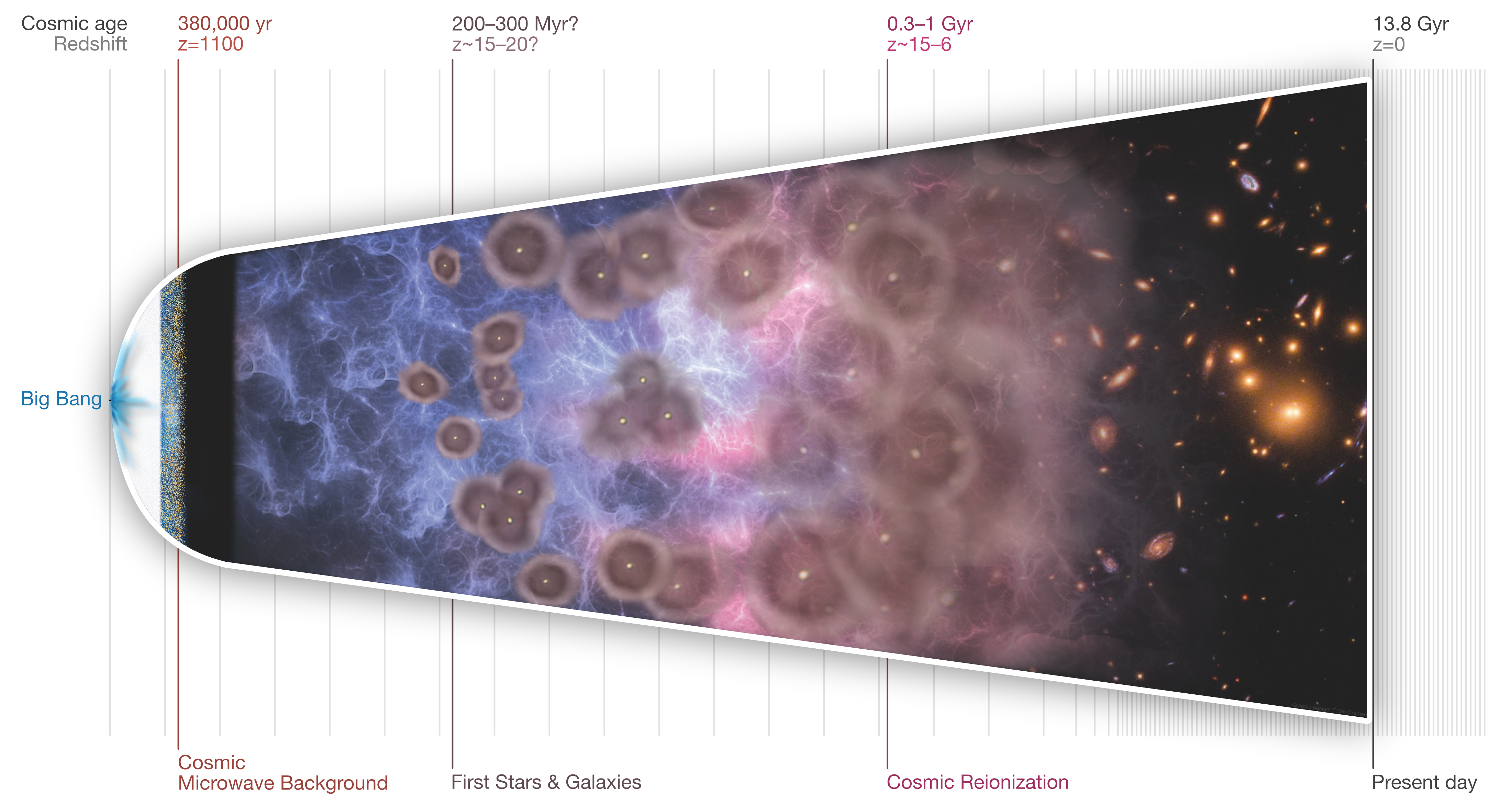}
    \caption{The cosmic timeline, from the origin of the Universe in the Big Bang, 13.8 billion years ago, till the present day. In the current standard picture,  the Universe underwent a period of accelerated expansion called ``inflation" that expanded the Universe by about 60 orders of magnitude. The Universe then kept cooling and expanding until the next major epoch of ``recombination'' about $4\times 10^5$~yr later when the first hydrogen atoms formed. This was followed by the ``Dark ages" of the Universe that lasted for a few hundred million years. The emergence of the earliest galaxies, a few hundred million years after the Big Bang, marked the start of the era of ``cosmic dawn". The first galaxies also produced the first photons capable of ionizing the neutral hydrogen atoms permeating space, starting the Epoch of Reionization (EoR), the last major phase transition in the Universe. In the initial stages of reionization, isolated galaxies (light yellow dots) produced ionized regions (gray patches) that grew and merged  until the Universe was fully reionized. Image Credit: DELPHI project (ERC 717001).}
    \label{fig:universe_timeline}
\end{figure*}

The first billion years of cosmic history were most eventful, rich with firsts and transformations (see a graphic representation in Figure~\ref{fig:universe_timeline}). After the Big Bang and initial accelerated expansion, the Universe's temperature and density constantly declined until photons and matter decoupled starting a period known as the cosmic ``Dark Ages". It is during this period that the seeds leading to the formation of the first stars and primordial galaxies were growing. Slow perturbations of dark matter accreted baryons until the conditions for the ignition of the first stars were met; the very first sources of light in the Universe. The energy, momentum, and enriched material ejected by these stars initiated the sparks that would eventually lead to the last major phase-transition of the Universe, the epoch of Reionization (EoR), consisting in the ionization of the neutral hydrogen permeating the Universe. This time of infancy for galaxies, where star formation processes and black hole (BH) formation and growth take place, is fundamental for our understanding of galaxy growth and evolution. Unfortunately, we know the least about this early time --- many questions remain about the basic facts and chronology.

When did the first stars form? Were they very different from the stars today?
Did any early galaxies have disks like our Milky Way, or were they disordered and bursting with star formation?
When did the first supermassive black holes form and fuel active galactic nuclei (AGN) powerful enough to impact the star formation of entire galaxies?
How quickly did supernovae enrich their surroundings with heavy elements and dust, the building blocks of planets and life?
And what was the timeline and topology of the last major phase transition of the Universe, as light steadily reionized the hydrogen gas permeating the Universe?

\JWST\ is now enabling us to address these questions with direct observations of the light of these early galaxies emitted more than 13 billion years ago and redshifted into the infrared by the expansion of the Universe. This is the first and overarching success that \JWST\ has unlocked: that it is possible to do these studies at all. With 100× greater sensitivity than previous telescopes at similar wavelengths \citep{Rigby2023}, \JWST's four science instruments NIRCam, NIRSpec, NIRISS, and MIRI can complete in hours of observations what previously would have taken years, and therefore been unfeasible.




We will refer to lookback time and distance in units of redshift ($z$), starting at $z = 5.6$, one billion years after the Big Bang 
according to standard cosmological models \citep{Planck18_cosmo},
and pushing back in time to the highest redshifts yet observed.


We dedicate this paper to the 20,000 people who spent decades to make \JWST\ an incredible discovery machine.




\section{The Physical Properties of Early Galaxies}\label{sec:properties}




\subsection{Extending the redshift frontier}
\begin{figure*}[h!]
    \centering
    \includegraphics[width=0.9\textwidth]{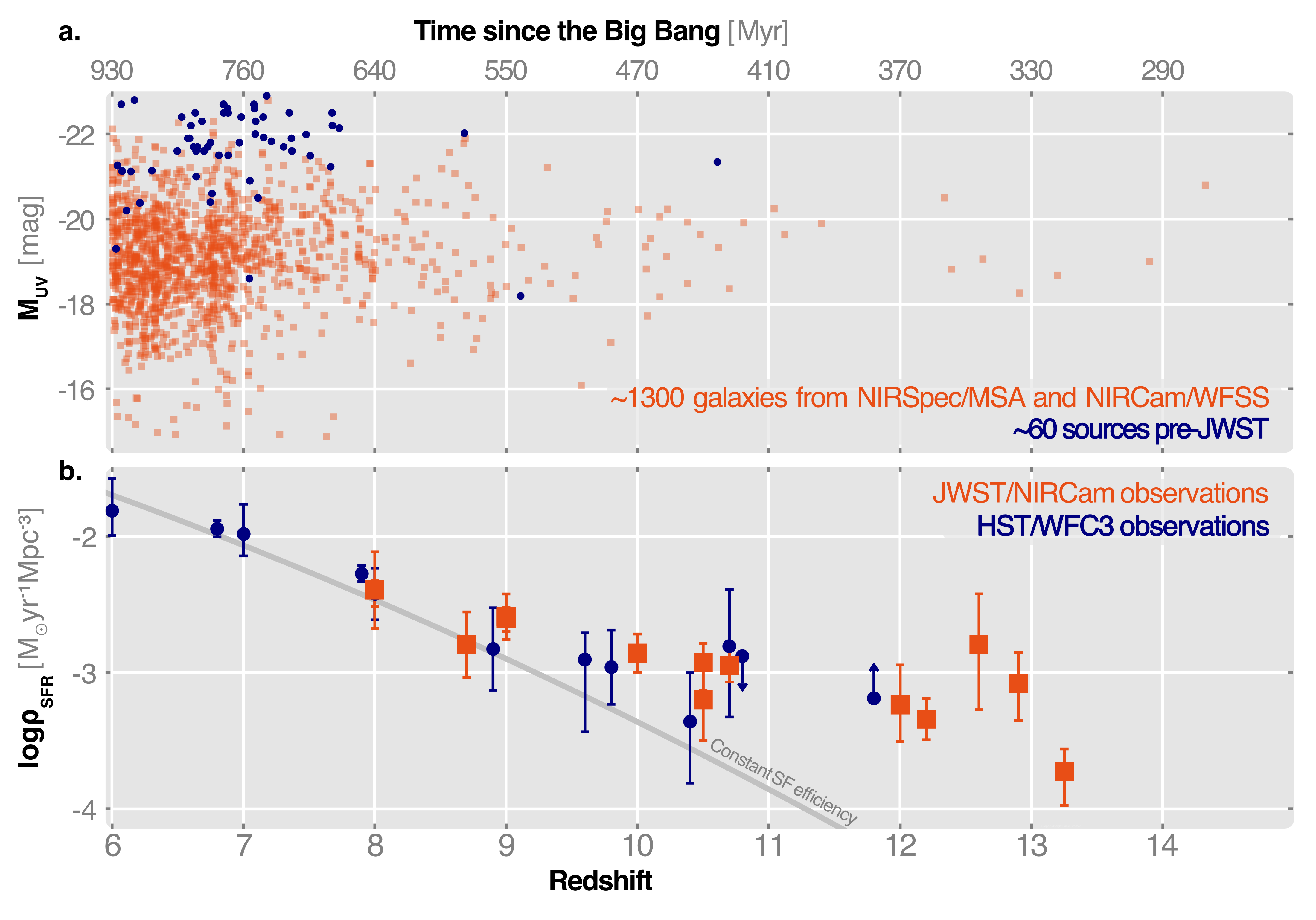}
    \caption{\textbf{a.} The distribution of absolute magnitudes and redshifts of spectroscopically-confirmed galaxies from pre-JWST candidates (blue dots) and from public JWST data sets (orange squares), showing the power of JWST to detect galaxies beyond redshift 6. The latter include compilations \citep{RobertsBorsani2024} and single targets \citep{castellano2024,carniani2024} observed with NIRSpec MSA observations, as well as NIRCam grism (FRESCO and EIGER; \citep{Oesch2023} and \citep{Daichi2023}, respectively). 
    \textbf{b.} The cosmic SFR density over the first billion years (adapted from Figure 17 of \citep{Harikane2024}, as seen from HST/WFC3 samples (dark circles), compared to JWST/NIRCam estimates (light squares). A model of constant star formation efficiency is plotted in grey, for comparison. The model and all literature points are derived from \citep{Harikane2024} (and references therein), where the latter are integrated down to $M_{\rm UV}=-18$ mag.}
    \label{fig:MUV-z}
\end{figure*}

{}
\JWST\ data reveal a robust population of high-redshift galaxies at $z>10$, beyond the reach of the Hubble Space Telescope (HST,  Figure~\ref{fig:MUV-z}). In most cases, the galaxies are selected through the signature, in multi-color imaging, of a sharp spectral cutoff blueward of the Lyman $\alpha$ line. This requires the galaxies to be reasonably luminous in the rest-frame UV. Many hundreds of such candidates have been identified at $z>8$ (see next Section),
and about twenty have been spectroscopically confirmed at $z>10$, with the current high mark at $z=14.32$ \citep{carniani2024}.
This record will surely not hold, as there is no technical limit on finding galaxies to $z=20$.









\subsection{UV Luminosity Function}{}


From these samples, \JWST\ has produced a precise determination of the galaxy UV luminosity function (LF) and luminosity density, $\rho_{\rm UV}(z)$, at redshifts $z \gtrsim 7$ (see Fig.~\ref{fig:MUV-z}, \citealp[e.g.,][]{Donnan2024,Bouwens2023a, Harikane2023a,PerezGonzalez2023b,McLeod2024,Robertson2023c,Willott2023}). The decrease of $\rho_{\rm UV}(z)$ is slower than generally expected before JWST, implying a more sustained early production of UV photons. Moreover, we observe that the bright end of the LF does not significantly evolve between $8 < z < 12$ \citep{Naidu2022b, Finkelstein2023b, Leung2023, Chemerynska2023} due to an unexpectedly large number of luminous ($M_{\rm UV} \simeq -21$) galaxies with blue spectral slopes $\beta < -2$ \citep{Topping2023, Cullen2023b}. The abundances are still limited in precision due to detected field-to-field variation attributed to large-scale structure \citep{Willott2023}, but these findings have important implications for cosmic reionization as discussed in Sec. \ref{sec:reionization}. 

While it has been proposed that these results might not support the standard version of the $\Lambda$CDM cosmology, they can likely be explained with a better understanding of star formation and baryon physics. 
Various physical scenarios, along with existing semi-analytic, semi-numerical models and simulations \citep[e.g.,][]{Wilkins2023c, mauerhofer2023}, have been proposed to explain the evolution of the LF. These include: (a) dust clearing by radiation-driven outflows \citep{Ferrara2023a, Ferrara2023b}, (b) star formation variability resulting in a flattening of the LF \citep{Mason2023,Mirocha2023,Pallottini2023}, (c) reduced feedback resulting in a higher star formation efficiency \citep{Dekel2023,Li2023e}, and (d) a top-heavy IMF \citep{Inayoshi2022}, although see \citep{cueto2023}.

The faint end of the LF is also important, as it depends on the nature of feedback processes acting in low-mass galaxies. Pre-JWST data suggested a steepening slope at higher redshift, but early JWST results appear to indicate a convergence to a logarithmic slope of $\alpha=-2$ in the LF, suggesting that extremely faint galaxies dominate the cosmic star-formation rate density. 


While the UV LF is based directly on observations, care must be taken in the interpretation. The UV LF contains several astrophysical degeneracies, as many processes can produce the same observed luminosity density. Viable models must simultaneously explain the tightly related stellar mass function evolution at lower redshifts \citep{Harvey2024, Weibel2024}.


\subsection{Stellar mass census} 

In contrast to HST, \JWST's wavelength coverage and spectral sampling of the rest-frame optical allows tracking of the bulk of the stellar mass, rather than solely the young stellar populations probed by UV light. These constraints are resolving earlier degeneracies regarding the contribution of emission lines, versus stellar continuum, enabling more accurate characterization of galaxy physical properties even beyond $z=10$ when the rest-frame optical emission of galaxies redshifts into the MIRI wavelength range.

The rest-frame optical observations enable improved constraints on stellar mass, placing the first robust constraints on the rate of mass buildup in early galaxies, back to the first 500 Myr \citep{Harvey2024, Weibel2024}. 
Pre-JWST observations with HST, Spitzer, and ALMA had hinted at the existence of rare, massive and red galaxies beyond $z>4$ that were typically excluded from rest-frame UV selections. 
JWST data reveals that these previously missed galaxies also contribute significantly to the stellar census \citep{Barrufet2023,Gottumukkala2023}. An exciting frontier is access to the rest-frame infrared provided by MIRI, which is crucial to prevent overestimates of stellar mass among massive, red, dust obscured sources \citep{Williams2023c, TWang2024}. 

Overly massive galaxy candidates emerged from the first \JWST\ data \citep[e.g.,][]{Labbe2023a,Xiao2023}. If real, these would require impossible or implausible star formation efficiencies \citep{BoylanKolchin2023}. To date, most follow-up observations of the earliest ($z>6$) massive candidates allow for alternative interpretations \citep{Desprez2023,Kocevski2023a}. Nonetheless, early massive quiescent galaxies have been spectroscopically characterized, requiring rapid formation at $z\sim10$  \citep{Carnall2023c,Glazebrook2023b,Nanayakkara2024, Setton2024}. The high number densities of these early quiescent galaxies, as implied by photometric samples, present severe challenges for theoretical galaxy formation models \citep[e.g.][]{Carnall2023b, PerezGonzalez2023a,Valentino2023}. Larger spectroscopic samples, which enable more detailed stellar population modeling, will further constrain their formation histories and test the extent of this tension.

Accurate characterisation of the optical rest-frame emission also paves the way to probe the growth of galaxies and the structural and kinematics of their stars and gas, and to map star formation sites and their feedback in terms of energy, momentum, metal enrichment injected in the galaxy medium, as discussed in detail below. 


\subsection{Star-Formation Histories}{}
Modeling the unprecedented broadband spectral energy distributions from \JWST\ reveals that $z>6$ galaxies contain stars that are young, with typical inferred ages of $\sim 50$~Myr, although younger ($<10$~Myr) \citep[e.g.,][]{Whitler2023,Whitler2024,Endsley2023a,Carnall2023a,Casey2023,PerezGonzalez2024a} and older \citep{Looser2023a,Strait2023} samples exist. Rest-frame optical ([OIII]+H$\beta$) emission-line equivalent width distributions and UV slopes also indicate bursty star formation, with both on and off modes \citep{Topping2022,Endsley2023b,Looser2023b}, also referred in the literature as stochastic star formation episodes.  
Theoretical predictions suggest that stochasticity can be induced by various feedback processes, such as (i) photoevaporation of molecular hydrogen, (ii) supernova explosions, and (iii) cosmological accretion/merging which dominate low-, intermediate-, and high-mass systems, respectively, and act on different timescales \citep{Pallottini2023, Mirocha2023, Sun2023b}. 
%
%
These results have fueled a resurgence of attention to outshining by the youngest stellar populations, including examples from spatially resolved data that can isolate multiple stellar populations \citep{GimenezArteaga2023,PerezGonzalez2023a,Fujimoto2024, Bradac2024}. Ultimately, as spectroscopic samples become more detailed and representative and analysis tools are optimized for the new datasets, we expect significant progress in this area.

\subsection{Structural and Kinematic Properties}{}
The dramatic improvement in the spatial resolution and wavelength coverage of \JWST's NIRCam instrument compared to its HST predecessors has enabled a sharper and more complete view of the distribution of stars in galaxies beyond $z\sim7$. The rest-frame optical size-mass relation has been measured into uncharted parameter space. At low masses, galaxies are more extended than conventional scaling relations would have predicted \citep[e.g.,][]{Cutler2023}. High-redshift galaxies at$z\gtrsim7$ roughly extend earlier relations, more rapidly evolving for quiescent galaxies than their star forming counterparts \citep[e.g.,][]{Yang2022,Morishita2023a,Ito2023}.

The same NIRCam imaging yields an unparalleled view of the stellar structures of those same galaxies. Although galaxies at 
high redshift ($z>3$) exhibit a broad diversity of morphologies, visual inspection has revealed that disk-like galaxies appear at early times ($\sim60\%$ at $z = 3$, $\sim30\%$ at $z\sim6-9$) \citep[e.g.,][]{Kartaltepe2023,Robertson2023a,Treu2023,Pandya2023}. These studies have included galaxies that were previously faint, or undetected in HST imaging \citep[e.g.,][]{Nelson2023a}. Remarkably, the imaging also enables the identification of more detailed structures, including galactic bars at $z\sim3$ \citep{Costantin2023b}, a disk plus proto-bulge at $z\sim7$ \citep{Baker2023}, and an enigmatic hint that some disk-like galaxies might actually be a population of elongated, or possibly prolate, systems  \citep{VegaFerrero2024,Pandya2023}. Machine-learning methods, informed by comparisons with simulations, begin to address the scalability of morphological classification \citep[e.g.,][]{VegaFerrero2024,HuertasCompany2023,Tohill2024}. Although challenges remain in identifying ongoing mergers from imaging alone, the merger fraction of sub-populations have been quantified out to $z\sim5-7$ \citep{Asada2024}.

Similarly, NIRSpec and the NIRCam grism have opened an unprecedented window into a range of kinematic structures, primarily from gas kinematics. In addition to a diversity of kinematic structures in smaller, Milky Way mass progenitors \citep[e.g.,][]{deGraaff2023,Fujimoto2024}, there are remarkable examples of massive rotating disks in the early universe \citep{Nelson2023a,Arribas2023}, and a host of non-virial structures in early galaxies driven by stellar and AGN-driven outflows observed in both emission and absorption \citep[e.g.,][]{Carniani2023,DEugenio2023b}.

\subsection{Clustered star formation, proto-globular clusters, single stars}
\begin{figure*}[h!]
    \includegraphics[width=0.9\textwidth]{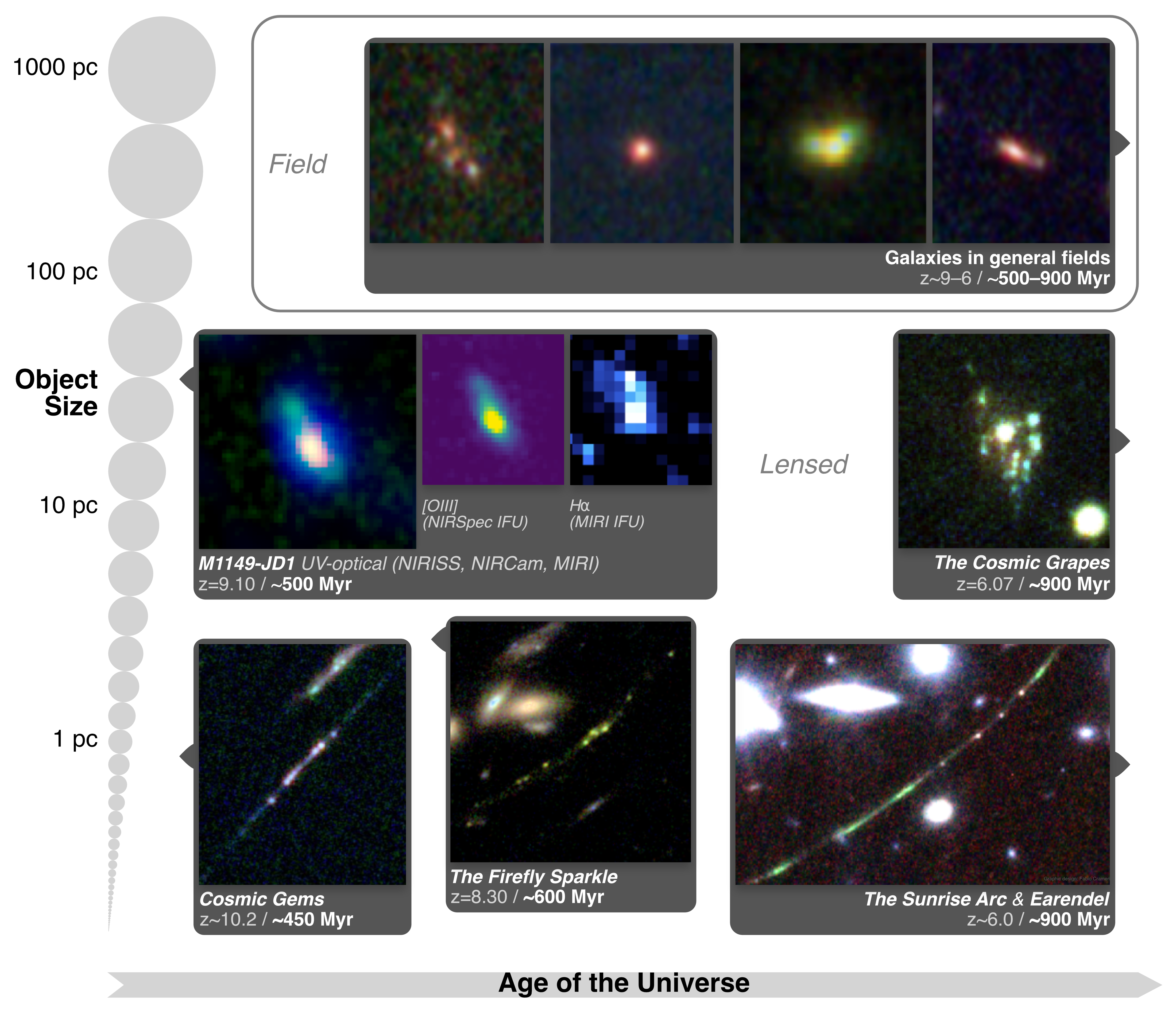}
    \caption{Resolved galaxy morphologies at redshift $> 6$ observed with JWST NIRCam (unless otherwise specified). Galaxies in the field (top row) show clumpy and dense structures \citep{Kartaltepe2023}. Thanks to gravitational lensing, the light from these compact galaxies is resolved into several stellar clumps down to scales of tens of parsecs  \citep[``The Cosmic Grapes";][]{Fujimoto2024}. In some cases, these clumps show strong emission lines as showcased for M1149-JD1 observed with NIRISS and NIRCam \citep{Bradac2024}, MIRI imaging and integral field spectroscopy \citep{AlvarezMarquez2023c}, and NIRSpec (GA-NIFS collab. in prep.) suggesting that intense episodes of star formation are concentrated within them.  Near the critical lines, the galaxy light is stretched into long arcs revealing bright compact  bound star clusters, with intrinsic sizes smaller than 10 parsecs \citep[``Cosmic Gems arc", ``Firefly Sparkle", ``Sunrise arc"][respectively]{Adamo2024, Mowla2024, Vanzella2023a} and single stars \citep[``Earendel"][]{Welch2022}. These stellar systems dominate the light of their galaxies, suggesting that star cluster might be a dominant star formation mode for young galaxies.}
    \label{fig:resolved_galaxies}
\end{figure*}

{}
\JWST's uniquely high spatial resolution has enabled us to map the physics of star formation within early galaxies (Figure~\ref{fig:resolved_galaxies}). Early galaxies in deep cosmic fields show  clumpy stellar structures and overall compact morphologies at scales of several hundred parsecs that dominate the galaxy's UV-blue optical light. Access to gravitational lensing enables us to resolve the galaxy light down to tens of parsecs or less. The clumpy structures observed in field galaxies break further down into smaller and denser stellar structures  \citep[e.g.,][]{Topping2024, Williams2023a, Hsiao2023a, Bradac2024, Claeyssens2023}. In some cases, these data reach parsec scale resolutions,  revealing young, massive star clusters in their interiors \citep{Vanzella2023a, Adamo2024, Mowla2024}. In general, more than 10\% of the total stellar mass observed in these galaxies is in star clusters, which dominate the FUV light of the host galaxies. The recovered ages show a progression, with the youngest clusters coincident with extreme emission line equivalent widths and elevated ionizing photon production rate per UV photons ($\xi_{ion}$). The latter is in agreement with predictions from stochastically populated  evolutionary models of massive star clusters, which show that clustered star formation overall increases $\xi_{ion}$ \citep{stanway2023}. 

Star clusters detected in early galaxies are consistent with being gravitationally bound, thus, potential proto-globular clusters (GCs) \citep{Vanzella2023a, Adamo2024, Mowla2024}. Their stellar densities are comparable or higher than GCs, enabling runaway stellar and BH mergers in their cores \citep{antonini2023}. Radiative and mechanical feedback from these proto-GCs is likely to be responsible for the extreme ionisation state of these early galaxies, as well as enhanced nitrogen abundances \citep{Cameron2023a, Topping2024}, which are potentially linked to the chemical pattern of multiple stellar populations found in Milky Way GCs \citep{gratton2019}. 
Star clusters within faint $z>6$  galaxies might be important units for reionization. \JWST\ observations of highly magnified galaxies will be fundamental in characterizing clustered star formation within these early galaxies, through cluster mass functions, cluster formation efficiency, and survival rates.    

Remarkably, gravitational lensing has also revealed individual stars as distant as Earendel at $z = 6$ \citep{Welch2024}.
JWST observations have begun revealing their properties, showing most are hot and luminous O and B stars \citep{Meena2023}.
Deeper JWST observations could deliver metallicities and outflow wind strengths of individual stars in the early universe \citep{Lundqvist2024}.

\subsection{Chemical Enrichment}{}

\begin{figure*}[h!]
    \centering
 \includegraphics[width=0.9\textwidth]{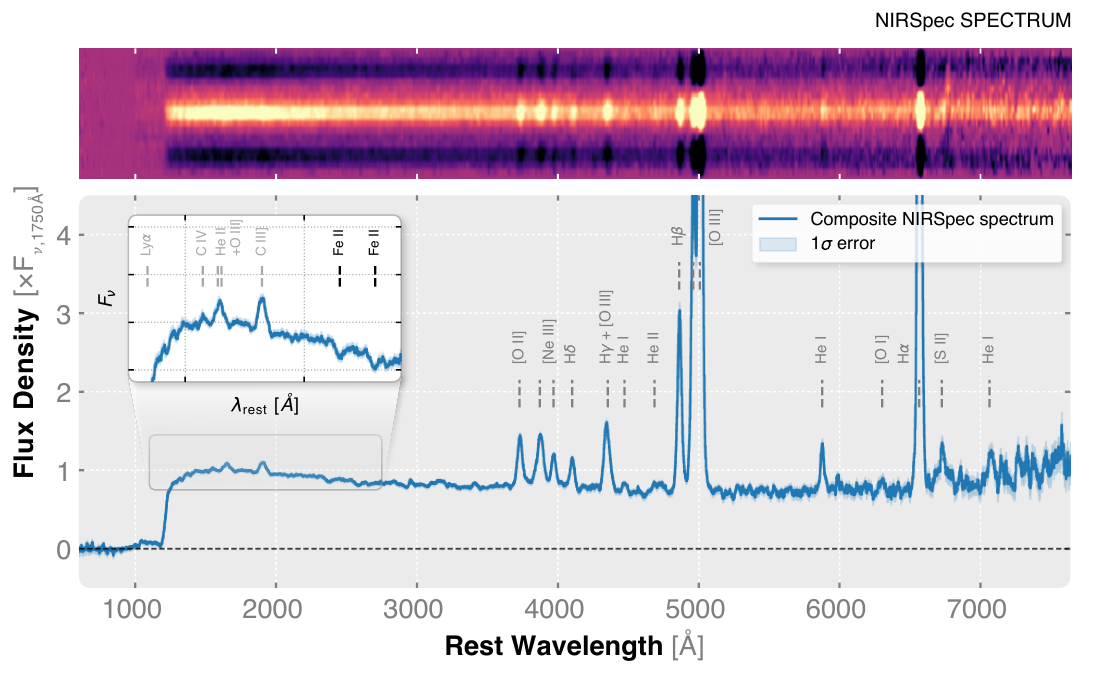}
    \caption{A composite spectrum using all publicly available $z\geqslant5$ low-resolution multi-object NIRSpec spectra. The unprecedented sensitivity and wavelength coverage of NIRSpec reveals a plethora of IGM, stellar, and ISM features, from Lyman and Balmer breaks to rest-frame UV and optical line emission. The plethora of features allow for characterizations of IGM opacity, stellar ages and masses, and gas-phase metallicities, to name a few. Plot adapted from Figure 3 of \citep{RobertsBorsani2024}.}
    \label{fig:spectra_stack}
\end{figure*}

\JWST's spectroscopic capabilities have illuminated the buildup of metals within the first billion years, where direct oxygen abundances of roughly 2--30\%\ solar have been measured for a dozen galaxies to date \citep{Schaerer2022,Nakajima2023,Laseter2024,Sanders2024}.
These ``gold standard" direct oxygen abundances are being used to anchor methods based on much brighter emission lines \citep{Sanders2023}, which are detected in large samples of galaxies out to $z\sim10$ (see Figure~\ref{fig:spectra_stack}).
To date, the highest-redshift measurements of N, C, O, Ne, Ar, and S abundances reach as far as $z=12.5$, only 350 Myr after the Big Bang \citep{DEugenio2023b}.

Dozens of robust oxygen abundances based on electron temperatures have been measured for the first time between $z\sim4-10$ \citep[see, e.g.,][for compilations]{Morishita2024a,Laseter2024}, spanning roughly 2--50\%\ solar, with the promise of many more to come. 
Early \JWST\ results demonstrate a continued gradual evolution towards lower total metallicity at fixed stellar mass beyond $z=4$, and intriguing evidence for redshift evolution in the baryon cycle at $z>8$ \citep[e.g.,][]{Nakajima2023,Curti2023b}.  
Importantly, \JWST\ has proven to be a powerful ``metal detector" for prominent high-ionization emission lines of C, N, and Ne \citep[e.g.,][]{Bunker2023a,Maiolino2023a,Chisholm2024,Topping2024}; these observations strongly suggest non-solar abundance patterns \citep[C/O and N/O; e.g.,][]{Cameron2023a,Jones2023c,DEugenio2023b} that may be the unique nucleosynthetic legacy of early stars.

To date with \JWST\, the lowest measured oxygen abundances reach a few percent of the solar value \citep[e.g.,][]{Atek2024,Maseda2023}, with evidence  reported for one object of abundances below 1\% \citep{Vanzella2023b}.
It remains unclear if this observed metallicity floor 
is the result of very rapid chemical enrichment on galactic scales (with associated light-weighted galaxy ages of $\lesssim$5 Myr) or a selection effect.
\JWST\ has thus not conclusively identified any first-generation galaxies that are composed entirely of pristine gas and Population III stars.  Nor have we yet seen other signatures of individual first stars, such as those that have been gravitationally lensed, or caught when they explode as the hypothesized pair--instability supernovae.

\subsection{Interstellar Dust}{}
\JWST\ has provided an unprecedented view of dust properties in early galaxies.
NIRCam photometry and NIRSpec spectroscopy have shown that the rest-UV continuum slopes of star-forming galaxies in the early Universe are typically blue, as is characteristic of young, dust-poor systems \citep[e.g.,][]{Topping2023,Cullen2023a,Cullen2023b,RobertsBorsani2024}. These results corroborate previous HST photometric studies of bright galaxies and, for the first time, extend our understanding to a more representative sample, even offering a tantalizing suggestion of dust-free stellar populations at $z>10$ \citep{CurtisLake2023}.
\JWST\ has also delivered the first constraints on nebular attenuation using hydrogen recombination lines past $z\sim 2.5$, revealing surprisingly little evolution at fixed stellar mass out to $z\sim7$ \citep{Shapley2023b,Sandles2023a}.
Given the general gas and metal evolution of galaxies, the constancy of attenuation over such a large redshift range is puzzling.

The incredible sensitivity of JWST 
has enabled the identification 
of optically faint, dusty sources that had previously eluded detection \citep[e.g.,][]{kokorev2023,Barrufet2023,PerezGonzalez2023a}. Such sources are now included,  
approaching a more complete census of star formation within the first billion years \citep{Xiao2023,Williams2023c}. 

The properties of dust grains reflect key processes of chemical enrichment, as well as heating and cooling in the interstellar medium. \JWST\ has enabled characterization of the composition and size distribution of the dust grains,  
with the detection of the 2175\AA\ feature at $z=6.7$, suggesting a rapid formation timescale for carbonaceous grains \citep{Witstok2023}. The extinction curve seems to flatten at high redshifts, perhaps reflecting a change of the dust grain size distribution, and of the main dust production sources \citep{Markov2024}.

\subsection{Large-scale Environments}{}
Early photometric and grism studies reveal substantially clustered megaparsec-scale distributions of galaxies and AGN beyond $z\gtrsim5$ \citep{Endsley2023a,HerardDemanche2023,Kashino2023,Wang2023a}.  A typical \JWST\ field contains multiple prominent peaks in the histogram of redshift, spanning over-densities $\delta=1-100$ on arcmin scales.  That high-redshift galaxies would be highly clustered was theorized, following mid-redshift measurements, but it indicates that care must be taken in interpreting the results of surveys in single sight-lines and narrow redshift intervals.  Substantial field-to-field variations are already observed \citep[e.g.][]{Eilers2024,Helton2024,willott2024}.

Further, at low redshift, it is well known that galaxy properties differ between high and low density environments.  It is not yet known what high-redshift properties will correlate substantially with environment.

\section{Active Galactic Nuclei (AGN) in early galaxies}\label{sec:AGN}

\begin{figure*}
    \centering
\includegraphics[width=0.9\textwidth]{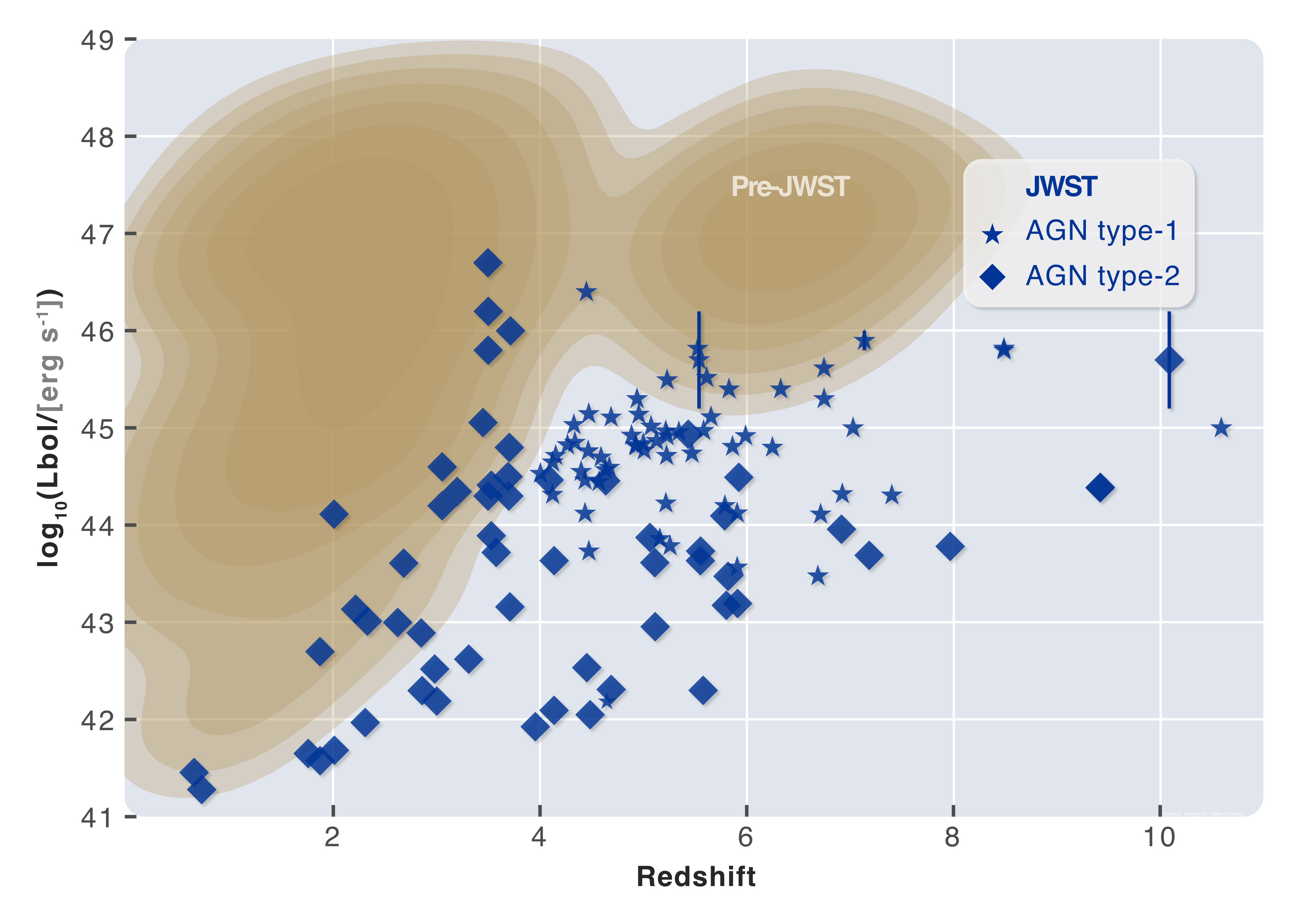}
    \caption{Distribution of bolometric luminosity of known AGN as a function of redshift, illustrating the discovery space opened by JWST at much lower luminosities and higher redshifts than probed by previous surveys (adapted from \cite{Scholtz2023b}). The brown shaded region shows the range of L$_{\rm bol}$ and redshift spanned by studies before JWST. The blue symbols show a compilation of AGN discovered by JWST, with stars showing type 1 (broad line) AGN and diamonds identifying type 2 (narrow line) AGN.}
     \label{fig:Lbol_z_AGN}
\end{figure*}

Before the advent of \JWST, observational studies were limited to detecting 
the most massive accreting BHs at high redshift. Such observations had confirmed the existence of extremely luminous and rare quasars ($L_{\rm bol}\geqslant 10^{45-46}\, \rm erg/s$, $\sim 1\, \rm Gpc^{-3}$) as early as $z\sim 7.6$, powered by BHs of $M_{\rm BH}\geqslant 10^{8}\, \rm M_{\odot}$. 
\JWST\ has not only detected,  for the first time, the stellar light of these quasars' host galaxies, but has also uncovered the presence of fainter AGN with $L_{\rm bol}\sim 10^{43-46}\,\rm erg/s$ (Fig.~\ref{fig:Lbol_z_AGN}). 
The demographics of BHs now encompass objects with masses of $\sim 10^6-10^9 \,\rm M_{\odot}$ at $z\sim 3-10$.

\subsection{Demographics and Identification}{}
Many candidate AGN have been identified through JWST imaging \citep[broadband, photometric SED,][and references therein]{Onoue2023,Barro2023,Labbe2023b,PerezGonzalez2024a,Williams2023c,Kokorev2024}.
Follow-up spectroscopy has confirmed the presence of AGN at high redshift, with many identifications occurring serendipitously during large spectroscopic surveys. AGN identifications primarily rely on the detection of a broad component of permitted lines, notably H$\alpha$, without any counterpart in the forbidden lines (e.g., [OIII]5008), ruling out outflows
\citep[e.g.][and references therein]{Kocevski2023b,Ubler2023b,Matthee2023b,Maiolino2023c, 
Greene2023}. 
The SEDs of some broad-line or candidate AGN have raised additional puzzles:  some are heavily obscured by dust \citep[$A_{\rm v}\sim1-4$,][]{Greene2023,Matthee2023b}, and some less extreme dust obscured AGN candidates have flatter-than-expected SEDs at rest-frame $\geqslant 1\mu$m. The latter might indicate that early obscured AGN could lack hot dust emission or that their rest-frame NIR is dominated by stars \citep[][with MIRI imaging]{Williams2023c, PerezGonzalez2024a}.

A large fraction of BH growth could occur in obscured phases;  there are now efforts to find these. 
While the classical BPT diagram seems to break down at high redshift, several narrow-line AGN were recently successfully identified from high ionization UV emission lines \citep[e.g.][]{Scholtz2023b,Chisholm2024}. Interestingly, 
most of these AGN could be X-ray weak 
\citep{Yue2024,Ananna2024,Maiolino2024Xray}, and various reasons are being explored (e.g., intrinsic weakness, extreme absorption).

\subsubsection{A large population}
These newly discovered AGN with $L_{\rm bol}\sim 10^{44}-10^{46}\, \rm erg/s$ appear more numerous than expected from the extrapolation of the luminous quasars' luminosity function, from X-ray selected AGN, or from 
some cosmological simulations \citep{Habouzit24}.
Specifically, number densities range from $10^{-5}$ to $10^{-3} \rm \, mag^{-1}\, cMpc^{-3}$ in the UV magnitude range $\rm -17<M_{\rm UV}<-21$ where $M_{\rm UV}$ is the magnitude of the AGN {\it and} their hosts

The fraction of early galaxies hosting an AGN
depends strongly on the sensitivity, AGN bolometric luminosity, 
and whether colour, size or spectroscopic criteria are used, such that the AGN fraction varies from $\sim$1\% to $\sim$20\%.

\subsubsection{BH masses}
JWST's
wavelength range has made it possible to estimate the BH mass of type-1 AGN at high redshift with the same tracers as those used in the local Universe, namely Balmer broad lines such as H$\alpha$ and H$\beta$. This finally enables consistent comparisons across redshift. For the small number of high-redshift quasars measured to date, the Balmer--derived masses are consistent with previously--derived Mg~II masses, within a scatter of 0.5 dex \citep[e.g.,][]{Yang2023b,Marshall2023,Bosman2023,Loiacono2024}.





\subsection{Quasar and AGN hosts}
JWST data have revealed, for the first time, the stellar light of a few quasar hosts at $z\geqslant 6$
\citep[e.g.,][]{Ding2023,Stone2023b,Yue2023}, 
indicating that quasars reside
in galaxies with stellar masses of $10^{10-11.5}\, \rm M_{\odot}$. 
In addition, JWST allowed 
astronomers to delve into the host properties (stellar mass, velocity dispersion) of the recently discovered AGN at $z\geqslant4$, and already raised intriguing questions.
Such AGN seem to be embedded in galaxies with stellar masses of $10^{8-10.5}\, \rm M_{\odot}$. Both these newly discovered AGN and high-redshift quasars have high BH-to-stellar mass ratios compared to those of the local Universe \citep{Ubler2023a,Harikane2023b,Maiolino2023a,Kokorev2023b,Juodzbalis2024}. However, selection effects
and uncertainties in both BH and stellar masses can play a role.


\subsection{AGN Large-scale environment}
On larger scales, JWST data have not only spectroscopically confirmed that 
quasars have close companions \citep{Wylezalek2022, Marshall2023}, but also demonstrated that some quasars are embedded in overdense environments, with an enhanced number of [O\textsc{iii}] emitters on Mpc scales
\citep{Wang2023a}. 
Ongoing JWST programs are now statistically evaluating whether quasars inhabit overdense, biased regions of the Universe. First results show that quasars live in diverse environments, including fields consistent with the average density of the Universe \citep{Eilers2024}. JWST will be key to disentangling whether this is due to a broad range of quasar dark matter haloes and/or the consequence of quasar feedback into their surroundings (e.g., suppression of nearby galaxy formation). 

\subsection{Implications for Black Hole Seeding and Growth}{}
Prior to \JWST, the existence of quasars at $z\geqslant 6$ suggested that their progenitors, referred to as BH seeds, 
must have formed with masses ranging from a few hundred solar masses (``light seeds'' formed from the remnants of PopIII stars) to about a million solar masses 
(``heavy seeds'') in the very early Universe. 
While \JWST\ has not ruled out 
any of the existing theoretical models, 
the confirmation of overmassive BHs (with high BH to stellar mass ratios) 
could hint towards the existence  of heavy seed channel(s). 
Those can 
form, e.g., from the collapse 
of high peak density fluctuations (``primordial BHs''), the collapse of super massive stars formed in atomic cooling haloes (``direct collapse'' seeds with $\sim 10^{3-5}\, \rm M_{\odot}$) or through hierarchical mergers of massive stars or 
black holes 
in compact stellar clusters (seeds with $\sim 10^{2-3}\, \rm M_{\odot}$).

Additional solutions to the abundance of  AGN 
and 
overmassive BHs include sustained super-Eddington accretion and 
BH mergers. 
Super-Eddington phases can account for a large fraction of BH growth under reduced BH feedback as shown in models \citep[][]{Inayoshi22,Schneider2023,Bennett2024} and supported by
recent observational findings  \citep{
Juodzbalis2024,Maiolino2023a}.
The contribution of BH mergers in their mass budget remains a longstanding question, tied to their mass, dynamics, and environment. 
Remarkably, \JWST\ has already revealed the existence of dual AGN at high redshift with separations of a few/several kpc \citep{Perna2023b,Ishikawa2024}; a population  surprisingly larger (by about an order of magnitude) than what is predicted by cosmological simulations. Additionally, there are growing indications of dual AGN on even smaller scales of a few 100~pc, which are likely in the process of merging 
\citep{Ubler2023b,Maiolino2023c}.

\section{Reionization}\label{sec:reionization}



The collective energy of the first stars, galaxies and accreting black holes -- including those too faint for even \JWST\ to detect directly -- transform the universe around them by heating and ionizing hydrogen (and later helium) gas in the intergalactic medium. This ``reionization" process sets the stage for all subsequent galaxy formation, as it impacts the ability of future dwarf galaxies to cool gas and form stars. We know that hydrogen reionization happened, but exactly when and \textit{how} it happened has been a major missing piece in our understanding of the first billion years.


Assuming reionization is driven by stars in early galaxies, the energy injection is commonly simply parameterised by the product of three terms. (1) The UV luminosity density of sources, $\rho_\mathrm{UV}(z)$, which describes the total abundance of high-redshift ionizing sources, as discussed above. (2) The production rate of hydrogen ionizing photons per UV photons, $\xi_\mathrm{ion}$, determined by the stellar populations. (3) The fraction of ionizing photons which escape the dense ISM of galaxies into the IGM, $f_\mathrm{esc}$. The rate of ionization is determined by the evolution of these quantities with redshift and galaxy properties, and the rate of hydrogen recombination in the IGM.


Before JWST we knew reionization ended approximately one billion years after the Big Bang ($z\sim5-6$), but not when it started. Due to the limited observations of the UV luminosity function for galaxies and AGNs, and local Universe indications that the escape fraction of ionizing photons from galaxies was modest ($<5$\%), the ionizing photon budget was tight. It was commonly assumed that reionization was accomplished by dwarf ($M_\mathrm{UV} \sim -13$) galaxies, by extrapolating a UV luminosity function steep at the faint end (slope $\sim-2$), and assuming escape fractions $10-20\%$. Due to the low observed number densities of UV-bright $z>6$ AGN, they were commonly assumed to not play a major role in hydrogen reionization.

\subsection{Sources of reionization}{}

With \JWST, we have made significant progress in understanding the reionization process. \JWST\ has shown that the ionizing photon budget is not necessarily tight and that high average values ($>20\%$) of the escape fraction are not necessary. As discussed above, we have now been able to measure the faint end of the UV LF at least up to $z\sim7-8$, and obtained information on the UV luminosity density of $M_\mathrm{UV} < -17$ sources within the first billion years. We have also obtained hints on the variability of star formation of these sources. We have found spectroscopically confirmed galaxies up to $z=13.2$, implying reionization may have started just a few hundred million years after the Big Bang. 

\JWST\ has for the first time spectroscopically measured the ionizing photon production of galaxies down to $M_\mathrm{UV}=-15.5$ \citep{Atek2024}. 
Direct measurements of $\xi_\mathrm{ion}$ from photometry and spectra of Balmer lines up to $z\sim6-9$ \citep[][]{PrietoLyon2023a,Endsley2023a,Simmonds2023,RobertsBorsani2024} find a range of values, $\log \xi_\mathrm{ion}\approx 25 - 26$ erg$^{-1}$ Hz, likely implying that galaxies have a range of star formation histories. Mean $\xi_\mathrm{ion}$ values are $\sim3\times$ higher than pre-JWST assumptions, reducing the need for high escape fractions.

\JWST's discovery and characterisation of faint AGN at $z>5$ (see above) has opened new discussions on the AGN contribution to reionization. However, as the majority of these high redshift AGN are likely UV-obscured or intrinsically faint, black hole accretion likely contributes no more than 25\% to ionizing budget \citep{Dayal2024a}.

There is still significant debate about the primary sources of reionization, in particular the contribution of faint galaxies ($M_\mathrm{UV} < -17$). This is driven by uncertainties in the galaxy properties including the ``burstiness'' of the star formation history, the IMF, and $f_\mathrm{esc}$. The foreground neutral IGM makes direct measurement of $f_\mathrm{esc}$ impossible within the epoch of reionization. Therefore, \JWST\ observations are being combined with those from ground-based telescopes and HST to identify galaxy properties that correlate with $f_\mathrm{esc}$ in the $z<3$ Universe, where direct observations of ionizing photons are possible. These correlations imply modest average $f_\mathrm{esc}$ ($\lesssim10$~\%) values in samples of tens of $z>6$  sources where direct comparisons with $z<3$ galaxies are possible \citep[e.g.,][]{Mascia2023a, Lin2024a}. 

Significant progress should be expected with future \JWST\ observations by surveying the faint ($M_\mathrm{UV} > -17$) galaxy population with deeper imaging and spectroscopy to understand their contribution to reionization.

\subsection{Morphology and growth of ionized regions}
\begin{figure*}[h!]
    \centering
\includegraphics[width=0.9\textwidth]{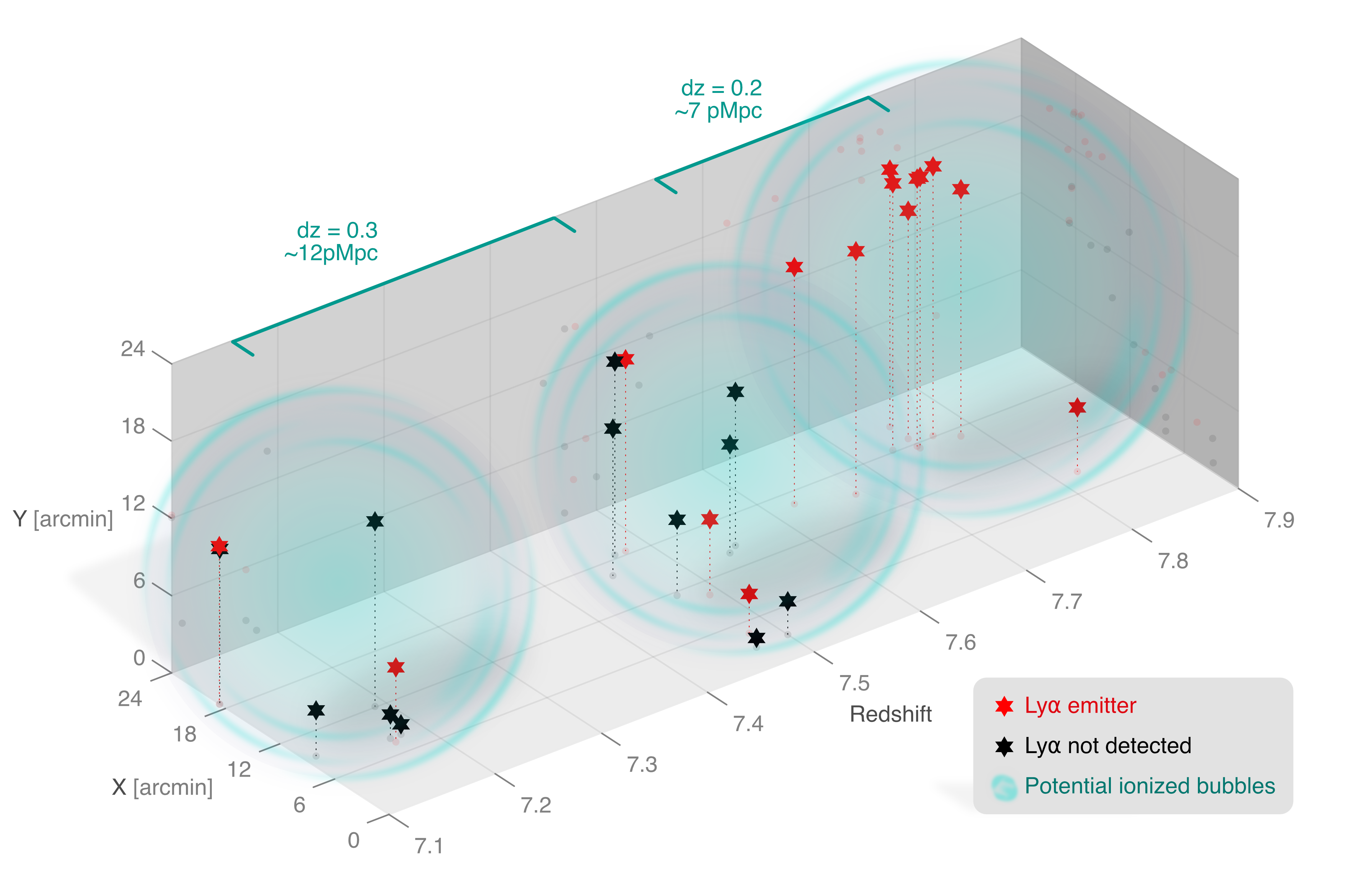}
    \caption{Spectroscopically-confirmed galaxies in the CEERS EGS field at $z = 7.1–7.8$, reproduced from \citep{Chen2024}. The presence of numerous Lyman-alpha emitting galaxies (red stars), including several with high equivalent widths ($>200$\,\AA) and Lyman-alpha escape fractions ($\gtrsim50\%$), in this field provides strong evidence for candidate ionized bubbles along the line of sight (shaded cyan regions -- for illustration purposes only). These early observations, primarily of $M_\mathrm{UV} < -19$ HST-selected sources, highlight the potential of JWST to create tomographic maps of ionized regions to learn about the reionization process on local scales.}
    \label{fig:bubble}
\end{figure*}

{}
\JWST\ is also enabling us to observe the reionization process directly. 
Reionization is predicted to be `patchy', with ionized bubbles growing around overdensities of galaxies (see Figure~\ref{fig:bubble}). The ionization state of the IGM can be traced by transmission of Lyman series photons -- either through the Lyman-$\alpha$ and $\beta$ forest, probing the very end stages of reionization at $z\sim6$, or Lyman-$\alpha$ emission from galaxies at all stages of reionization. \JWST\ has shown that the $z\sim6$ Lyman-$\alpha$ and $\beta$ forest transmission measured in quasar spectra is correlated with distance from galaxies, suggesting that galaxies (and their overdensities) dominated the ionizing emissivity at $z\sim6$ \citep{Kashino2023}.
A downturn in the average Lyman-$\alpha$ emission from galaxies at $z\gtrsim6$ has been confirmed \citep{Nakane2023,Chen2023b}, implying we are observing a predominantly neutral universe at higher redshifts. \JWST\ spectroscopy enables us to infer the ionization state of the IGM at $z>10$ \citep{Umeda2023} hence providing an independent constraint on the total star formation rate density at early times, probing sources below the detection limits of even JWST.

\JWST\ also finds strong evidence for ionized regions, giving the first opportunity to understand how reionization is occurring on local scales. Early results have shown large field-to-field variations in {Lyman-$\alpha$} emission visibility, even in overdensities \citep{Morishita2023b,Chen2024,Napolitano2024}, and also highlighted the importance of understanding {Lyman-$\alpha$} absorption by dense residual neutral gas within and around galaxies \citep{Hsiao2023a,Heintz2023b,Chen2024}.
The sizes of ionized regions are predicted to be determined by the clustering strength, luminosity and star formation histories of the dominant reionizing sources \citep[e.g.,][]{Lu2024} and can be probed directly via the spatial variations in {Lyman-$\alpha$} transmission. 

\JWST\ has enabled the first measurements of {Lyman-$\alpha$} escape fractions (from Balmer lines), robust fluxes and velocity offsets from systemic in $z>7$ galaxies. The discovery of sources with high Lyman-$\alpha$ rest-frame equivalent width and Lyman-$\alpha$ escape fraction, and low velocity offsets, in overdense regions implies these galaxies reside in ionized regions $\gtrsim1$\,pMpc \citep{Saxena2023a,Tang2024,Chen2024}. By spectroscopically confirming galaxies in these regions, we are starting to build 3D maps \citep[Figure~\ref{fig:bubble},][]{Chen2024,Witstok2024} which probe the contribution to reionization from faint sources below \JWST's detection limits \citep{Whitler2024}.

Future systematic spectroscopic studies of Lyman-$\alpha$ visibility as a function of environment and galaxy properties will help us understand how ionized regions grew. This will enable us to constrain ionizing photon escape fractions and star formation in sources beyond \JWST’s detection limits.

\section{Conclusions}
\label{sec:conclusions}
\JWST's sharp imaging, broad wavelength coverage, exquisite spectroscopy, and two-order-of-magnitude improvement in sensitivity have unlocked the study of how galaxies evolved in the first billion years. The high--redshift Universe has emerged in stunning definition and the initial discoveries of \JWST\ are rewriting astronomy textbooks.

\JWST\ has revealed that galaxies and black holes assembled, formed significant stellar mass, dust, and synthesized the elements of the periodic table much earlier than was expected. The remarkable abundance of luminous galaxies at such early times hold the promise of upending theories of primordial galaxy formation. The discovery of very dense and massive star clusters urge us to decipher the role that these tiny stellar systems play in the evolution of early galaxies. The large numbers and apparent high masses of accreting supermassive black holes found by \JWST\ have renewed interest in channels that generate intermediate--mass seeds. \JWST\ has directly measured the abundance of chemical elements in these early galaxies, finding low ($ \gtrsim  2\%$) but non--zero oxygen abundance, which indicates either that the Universe rapidly self--enriched, or that \JWST\ has not yet reached the epoch of the very first galaxies.  Whereas before \JWST\ we knew that the Universe must have reionized, but did not know which sources were responsible, \JWST\ has now found thousands of galaxies in the epoch of reionization, and associated individual ionized bubbles with star-forming galaxies, beginning to map out the topology of reionization.  

We are gathering many pieces in the puzzle that describes the chronology of the first billion years of the universe, however, the emerging picture is still scattered. In the coming years, as many more discoveries are made, we will be able to connect these subfields, and comprehensively describe the first billion years of cosmic history.



\clearpage

\section{Acknowledgments}
The authors wish to thank ISSI for sponsoring the 2024 Breakthrough Workshop, and the ISSI staff for their wonderful welcome and support. The authors are grateful to their collaborators, who made this paper possible. Collectively, we are grateful to the large community of scientists and engineers, worldwide, who designed, built and commissioned JWST and made a decades long astronomer dream a reality.

\noindent {\bf Author Contributions:} All of the authors were invited participants in a Breakthrough Workshop at the International Space Science Institute, titled ``The Chronology of the very early Universe according to JWST'', which was held 11--15 March, 2024, in Bern, Switzerland. Most participants were successful JWST Cycle 1 PIs.  All authors actively participated in workshop discussions and writing of the paper.  
The conveners of the workshop were Antonella Nota, Angela Adamo, Gabriel Brammer, Dan Coe, Pascal Oesch, and Jane Rigby.  
The editors were the conveners and Daniel Eisenstein. 
The leads of the individual sections of the paper were  
Pascal Oesch, Dan Coe, Angela Adamo, Rachel Bezanson, Daniel Eisenstein, Andrea Ferrara, 
Melanie Habouzit, Roberto Maiolino, 
Charlotte Mason, Alice Shapley, and Massimo Stiavelli.
The discussion moderators were  
Jane Rigby, Danielle Berg, John Chisholm, Pratika Dayal, Ivo Labbe
Michael Maseda, Jorryt Matthee, Kristen McQuinn, Dan Stark, Allison Strom, Christina Williams, and Dominika Wylezalek. 
The discussion notetakers were Gabe Brammer, Anna de Graaff, Taylor Hutchison,  Jeyhan Kartaltepe, Rui Marques and Rohan Naidu. The authors are grateful to Mark Dickinson for a careful read of the final manuscript, and to Fabio Crameri  (ISSI) for his expert help designing the very best figures.   
%








\bibliography{JWST-galaxies}{}
\bibliographystyle{aasjournal}


\end{document}